# Implementation of a speckle correlation based optical lever (SC-OptLev) with extended dynamic range

A. VIJAYAKUMAR,[1,*] D. JAYAVEL,[1] M. MUTHAIAH,[1] SHANTI BHATTACHARYA,[1] AND JOSEPH ROSEN[2]

[1]*Department of Electrical Engineering, Indian Institute of Technology Madras, Chennai – 600036, India.*
[2]*Department of Electrical and Computer Engineering, Ben Gurion University of the Negev, P.O. Box 653, Beer-Sheva 8410501, Israel.*
*\*physics.vijay@gmail.com*

**Abstract:** A speckle correlation based optical lever (SC-OptLev) is constructed for the measurement of small changes in the angle of orientation of a surface. The dynamic range of SC-OptLev is found to be twice that of a conventional OptLev for the same experimental configurations. Different filtering mechanisms are implemented and the correlation results are compared. Two types of computer automated SC-OptLevs, open source based computing system with a low-cost image sensor and a commercial computing system, are presented with assistive computational modules.

## 1. Introduction

In 1826, Poggendorf invented the Optical Lever (OptLev) for improving the sensitivity of theodolites, which was later adapted by Gauss and Weber in 1846 for their experiments [1]. The OptLev is now widely used for the measurement of small changes in the angle of orientation of a surface with a high accuracy [2-5]. OptLevs are also part of the Laser Interferometer Gravitational-Wave Observatory (LIGO) and Kamioka Gravitational Wave Detector (KAGRA), where small mirror tilts are monitored and corrected in real-time [6-8]. The OptLevs in LIGO uses quadrant photodiodes (QPD) to detect the tilts occurring due to the radiation pressure and generate appropriate control signals to correct the misalignment. Even though the OptLevs with QPD are useful for measuring the variation in the angular orientation, there are several limitations owing to the inability to detect other variations. Speckle correlation based sensing, on the other hand, can detect almost any variation in the surface ranging from local variations to bulk movements [9]. Speckle correlation based OptLev (SC-OptLev) was introduced by Hinsch *et. al.* using a VanderLugt 4F correlator with a matched filter for non-destructively monitoring structural changes in a surface [9]. Later, digital speckle correlation was implemented for a variety of surface studies [10-17].

 In a conventional OptLev [18], light from a coherent source is incident on a reflective surface whose angular orientation is to be measured. The light deflected from the reflective surface is incident on an image sensor and the variation in the angular orientation of the surface is measured from the shift in the location of the spot as shown in Fig. 1(a). In this manuscript, we propose and demonstrate a simple, low-cost SC-OptLev for high precision measurements of small changes in the angle of orientation of a surface. A dynamic range double that of the conventional OptLev is realized in the proposed optical configuration. Different decorrelation techniques ranging from the well-known Lucy-Richardson iterative algorithm [19] to the most recently developed non-linear correlation [20] were implemented in the SC-OptLev and their performances were compared. One of the main drawbacks associated with SC-OptLevs compared to conventional OptLevs is that it is difficult to monitor the variations in the angle of orientation of a surface in real-time. In order to overcome this difficulty, we propose and demonstrate computational modules for real-time monitoring of changes in angular orientation. Two different types of computational optical modules are presented. The first module uses a low-cost web camera and completely open source based computational environment for regular measurements. The second module uses a scientific camera and a commercial computational environment for more sophisticated measurements.

The manuscript consists five sections. In the second section, the principle of operation and methodology of an OptLev is discussed. The experimental results are compared in the third section. In the fourth section, the computational framework for automation using Ubuntu-Octave and Windows-MATLAB are presented with assistive supplementary files. The summary of the technique and the future perspectives are presented in the final section.

## 2. Methodology

The optical configurations of the conventional OptLev and the proposed lensless SC-OptLev are shown in Figs. 1(a) and 1(b), respectively. In Fig. 1(a), light from a coherent source illuminates a plane mirror at an angle of 45° with respect to the optical axis and the reflected light is incident on an image sensor located at a distance of $L$ from the mirror. When the plane mirror is rotated by an angle of $\beta$, the optical beam is deviated by an angle $\theta = 2\beta$ and the beam shifts by a distance $d = L \cdot \tan(\theta)$ on the image sensor. As the parameters $L$ and $d$ are directly proportional, the angular sensitivity of the measurement can be increased by increasing $L$. The above relationship enables the conventional OptLev to measure small variations in the orientation of objects with a high accuracy. For an image sensor with $n \times n$ pixels and a pixel size of $\Delta$, the size of the image sensor is $\Delta \cdot n \times \Delta \cdot n$. Assuming a 1-D angular variation, the dynamic range of measurement of $\beta$ under small angle approximation is $R = \pm \Delta \cdot n / 4L$. The sensitivity defined as the minimum angle for which the detected signal is shifted by a full pixel is $S = (\Delta / 2L)$. From the above relationships, a trade-off between the dynamic range and the sensitivity is seen. When $L$ is increased, the sensitivity is improved ($S$ becomes smaller), while the dynamic range decreases and vice versa.

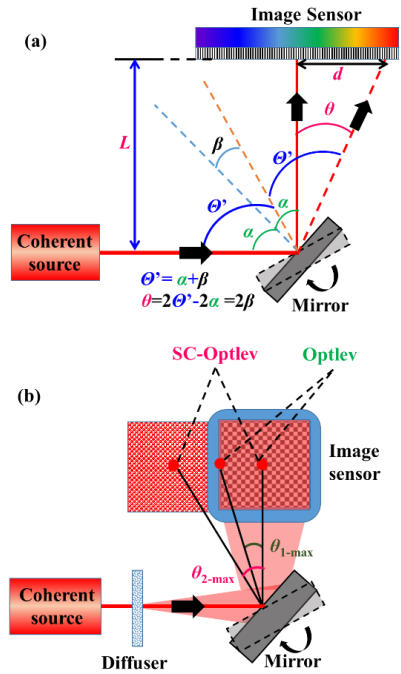

Fig. 1 Optical configuration of (a) conventional OptLev and (b) SC-OptLev.

The optical configuration of SC-OptLev is shown in Fig. 1(b). Light from a coherent source is modulated by a diffuser plate, such that a speckle pattern is reflected from a plane mirror and collected by the image sensor. This speckle pattern is an optical signature unique for each wavelength of the light, for the phase function of the diffuser, for the distance $L$ and for the angle of orientation of the mirror $\beta$. Any change in the above parameters will result in a different speckle pattern, but only a change in $\beta$ shifts the speckle pattern by a distance $L \cdot \tan(\theta)$. Therefore, by recording the initial speckle pattern and by monitoring the variation in the speckle pattern in real-time, the measurement of the angle orientation of the tested mirror can be simplified to a pattern recognition task [21]. Assuming that the illuminating source is frequency stabilized, the location of the diffuser and the distance $L$ are constants, any change in the speckle pattern can be related to the change in the angle $\beta$ of the mirror orientation. The light reflected from the mirror is tilted, where the tangents of the tilt is given as $[\tan(\theta_x), \tan(\theta_y)] = (d_x/L, d_y/L)$. By correlating the initial speckle pattern $I_0$ with the new speckle pattern $I_\beta$ after tilting the mirror, the location of the correlation peak can be detected from the pattern $C = |\mathfrak{F}^{-1}\{|\tilde{I}_0|exp[i\,arg(\tilde{I}_0)]|\tilde{I}_\beta|exp[-i\,arg(\tilde{I}_\beta)]\}|$, where $\tilde{I}_0$ and $\tilde{I}_\beta$ are the Fourier transform of $I_0$ and $I_\beta$, respectively. Measuring the displacement $d = (d_x, d_y)$ of the correlation peak from the origin, which is the autocorrelation peak's location for $I_0$, the 2-D angular variation can be estimated as $\beta_x = \tan^{-1}(d_x/L)/2$ and $\beta_y = \tan^{-1}(d_y/L)/2$. If the speckle pattern on the image sensor is much smaller than the area of the image sensor then there is no advantage of SC-OptLev compared to conventional OptLev. However, if the area of the speckle pattern is at least equal to the size of the area of the image sensor, then SC-OptLev can exhibit a dynamic range, which is twice that of a conventional OptLev as compared in Fig. 1(b).

In the previous studies [9-11], a matched filter had been implemented to obtain the location of the correlation peak. However, studies conducted later revealed that the matched filter is not the ideal candidate for a decorrelation, as it suffers from substantial background noise. In this study, the decorrelation has been studied using a variety of filters and techniques, namely a matched filter, a phase-only filter [21], a non-linear correlation [20], a Wiener filter [22] and the Lucy-Richardson iterative algorithm [22]. The results are compared based on the background noise, computation time and the width of the correlation peak. As much as the correlation peak becomes narrower, its location accuracy is improved. In the phase-only filter technique, the correlation is computed as $C = |\mathfrak{F}^{-1}\{exp[i\,arg(\tilde{I}_0)]|\tilde{I}_\beta|exp[-i\,arg(\tilde{I}_\beta)]\}|$, which suppresses the background noise and narrows the correlation peak [21]. In the case of the non-linear correlation, the correlation is computed as $C = |\mathfrak{F}^{-1}\{|\tilde{I}_0|^r exp[i\,arg(\tilde{I}_0)]|\tilde{I}_\beta|^o exp[-i\,arg(\tilde{I}_\beta)]\}|$, the values of $o$ and $r$ are tuned to minimize the entropy $S(o,r) = -\sum\sum \phi(m,n)log[(m,n)]$, where $\phi(m,n) = |C(m,n)|/\sum_M \sum_N |C(m,n)|$ [20], and $(m,n)$ are the indexes of the correlation matrix. The search for the optimal parameters of $o$ and $r$ is done only once for some arbitrary tilt and the same parameters are used for the entire measurements. Wiener filter behaves like an ideal inverse filter ($o = 1$, $r = -1$), when there is no background noise, and requires the estimation of the noise in order to deconvolve noisy distributions.

### 3. Experiments

An experimental setup was built as shown in Fig. 1(b). Light from a He-Ne laser ($\lambda = 632.8$ *nm*) illuminated a diffuser with 1000 grit polish. The light diffracted from the diffuser was incident on a plane mirror oriented at an angle of 45° with respect to the optical axis. The light diffracted from the mirror was captured by Thorlabs Camera DCC1240M (1024×768 pixels, pixel size: 4.65 *μm*). The distance between the mirror and the camera was 7 *cm*, while the distance between diffuser and mirror was 3 *cm*. Nearly 1/3rd of the area of the image sensor in the central part was used for the demonstration i.e., about 340×256 pixels. The maximum tilt allowed in a conventional OptLev is ($\beta_{x\text{-max}} = 0.32°$, $\beta_{y\text{-max}} = 0.24°$). The mirror was tilted along the $x$ and $y$ directions independently. Three cases were tested, a tilt only along the $x$ direction, only along *y*, both beyond the limit of the image sensor and the third case was a tilt along *x-y* direction within the limit of the image sensor. The images of the speckle patterns $I_0$, $I_x$, $I_y$ and $I_{xy}$ are shown in Figs. 2(a)-2(d), respectively. The four images were zero-padded to be twice their initial lengths and four cases were considered: autocorrelation of $I_0$, cross-correlations of $I_0$ with $I_x$, $I_y$ and $I_{xy}$. The correlation results using matched filter, phase-only filter and non-linear correlation for ($o$ = -0.3, $r$ = 0.8) are shown in Figs. 2(e)-2(h), 2(i)-2(l) and 2(m)-2(p), respectively. The correlation result using Wiener filter is shown in Figs. 2(q)-2(t) and the results of the Lucy-Richardson method are shown in Figs. 2(u)-2(x). The Octave code, for the matched filter, phase-only filter and non-linear correlation, is given in Code–1. The MATLAB code, for the Wiener and Lucy-Richardson Filters, is given in Code–2.

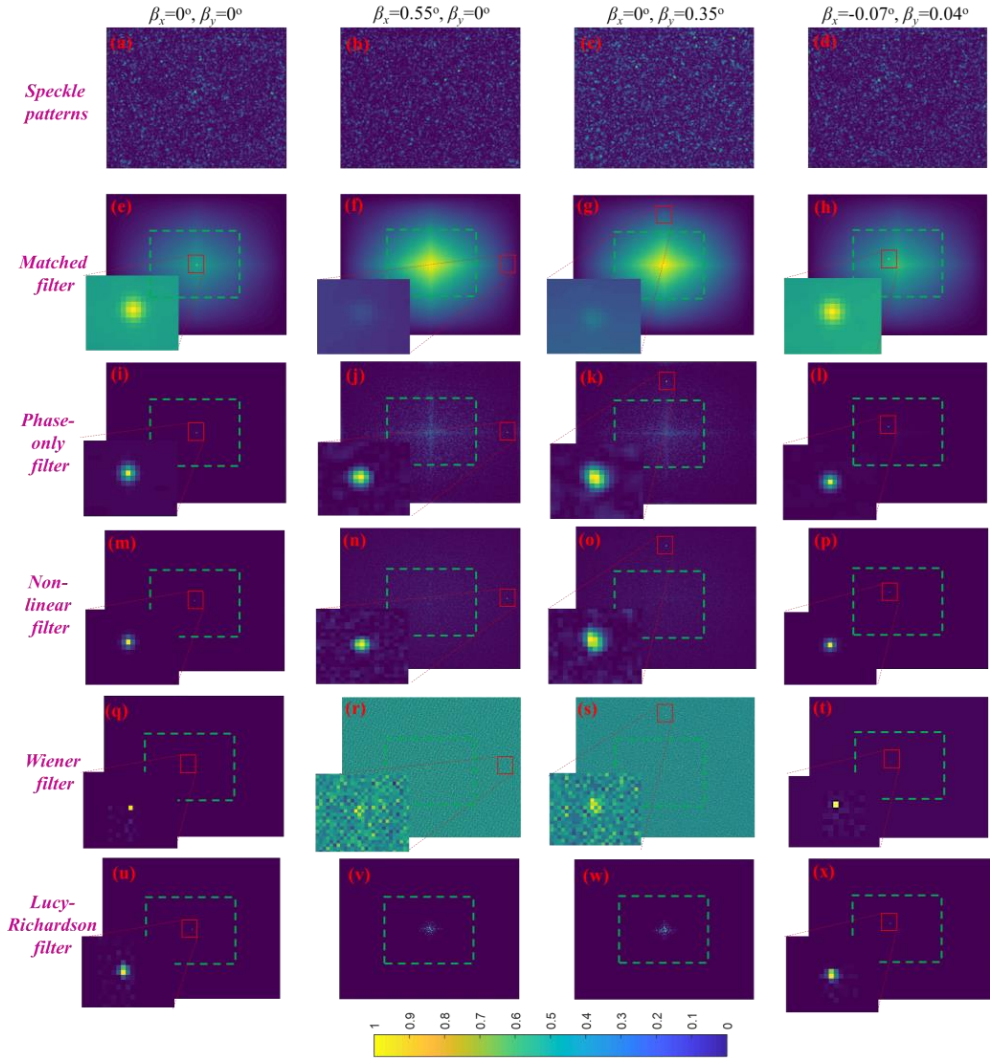

Fig. 2 (a)-(d) Speckle patterns, correlation results using (e)-(h) matched filter, (i)-(l) phase-only filter, (m)-(p) non-linear filter, (q)-(t) Wiener filter and (u)-(x) Lucy-Richardson filter for ($\beta_x$=0°, $\beta_y$=0°), $\beta_x$=0.55°, $\beta_y$=0.35° and ($\beta_x$=0.07°, $\beta_y$=0.04°), respectively. Green dotted line indicates the sensor area.

The shift of the correlation peaks were noted to be 289 and 186 pixels which corresponds to ($\beta_x$ = 0.55°, $\beta_y$ = 0°) and ($\beta_x$ = 0°, $\beta_y$ = 0.35°), respectively for the first two tilts of the mirror and a shift of -36 pixels and 23 pixels along the *x* and *y* directions with ($\beta_x$ = -0.07°, $\beta_y$ = 0.04°) for the final tilt. From the correlation results, it is seen that the background noise is increased with the increase in the variation in the angular orientation. The results from Wiener filter and Lucy-Richardson method have a sharper correlation peak with less background noise for smaller changes in the angular orientation while both methods fail for larger variations. Moreover, Lucy-Richardson method is not suitable for real-time monitoring as it is an iterative process and consumes more computational time compared to the other methods especially if the number of required iterations is large. By comparing the results, it is clear that the non-linear correlation has the minimum background noise and narrowest correlation peak for all types of angular variations due to its adaptive nature [20]. The execution time of the above methods are almost equal except that a longer time was noticed for the Lucy-Richardson method owing to its iterative nature. The FWHM of the correlation peak, the signal to noise ratio (SNR) measured as the inverse of the background noise and the computation time obtained using '*tic*' and '*toc*' at the beginning and end of the program (Octave/MATLAB) are given in Table 1 for the above methods and for the case of *x-y* tilt. From this comparison, it is found that the non-linear correlation is a better choice compared to the other evaluated filters and methods of the SC-OptLev.

Table 1. Comparison of SNR, FWHM and computation time for Matched filter, Phase-only filter, Non-linear correlation, Wiener Filter and Lucy Richardson Method for ($\beta_x$=0.07°, $\beta_y$=0.04°) (CT-Computation time, FWHM – Full Width at Half Maximum, SNR – Signal to Noise Ratio and NA – Not Applicable as no signal was seen).

| Method | $\beta_x$=-0.07°, $\beta_y$=0.04° | | $\beta_x$=0.55°, $\beta_y$=0°  | | |
|---|---|---|---|---|---|
| | SNR | FWHM (µm) | SNR | FWHM (µm) | CT (*s*) |
| Matched-Filter | 1.4 | 28 | 1 | 28 | 0.16 |
| Phase-only Filter | 40 | 9.3 | 6.5 | 18.6 | 0.18 |
| Non-linear Filter | 258 | 9.3 | 18.5 | 18.6 | 0.29 |
| Wiener Filter | 41 | 2.3 | 1.8 | 18.6 | 0.29 |
| Lucy Richardson Method | 89.8 | 11.6 | NA | NA | 0.78 |

### 4. Software structure for automation

A drawback of SC-OptLev is that the real-time monitoring of the angular variations is more challenging than in the conventional OptLev. In the conventional OptLev, the change of the angular orientation can be monitored by viewing the streaming video, which shows the shift in the location of the light spot in real-time. On the other hand, in SC-OptLev, the streaming video shows a speckle pattern in motion, from which it is not possible to ascertain the location of the correlation peak or the angular deviation.

Most of the manufacturers of scientific cameras provide interface software in MATLAB and/or LabVIEW. The cost of such scientific cameras is about 1000$. Commercial operating systems such as Windows (~100$) and MATLAB/LabVIEW packages (~500$), increase the overall cost of what is essentially a simple experiment. In this section, we present a completely open-source based computational module for real-time monitoring of the angular deviation using a web camera (~20$) for SC-OptLev. As a result, the cost can be substantially reduced from the above ~1600$ to ~20$. The software structure using Debian GNU/Linux (Ubuntu 16.04)-GNU Octave (4.3.0) [23,24] grabs a video and processes it frame by frame by implementing the non-linear correlation in real-time and finally the program displays the motion of the correlation peak in a virtual window. A peak detector function is incorporated to know the location of the correlation peak. From this data, in addition to the initial location of the autocorrelation peak obtained for $I_0$ and the distance *L*, the angular deviation along *x* and *y* directions can be calculated by a simple trigonometry.

The schematic of the optics integrated with the computer module is shown in Fig. 3. The process sequence is as follows: (a) Grab a frame $I_0(t = 0)$ from the web camera at time $t = 0$ and store it as a matrix/image, (b) start real-time image acquisition, (c) grab a frame $I_\beta(t)$ and apply the Non Linear Correlation Function on $I_0(t = 0)$ and $I_\beta(t)$, (d) display the result real-time and (e) detect peak and display angle ($\beta_x$, $\beta_y$). The code for automation and the details for building dependencies are given in Code-3. The peak detection and angle calculation code is given in Code-4. An alternative version for MATLAB users is given in Code-5. The mirror in the SC-OptLev was tilted and the motion of the correlation peak real-time was recorded as a video for two cases: Without zero-padding and with zero-padding as shown in visualization–1 and visualization–2, respectively. The video recorded using MATLAB is shown in visualization–3.

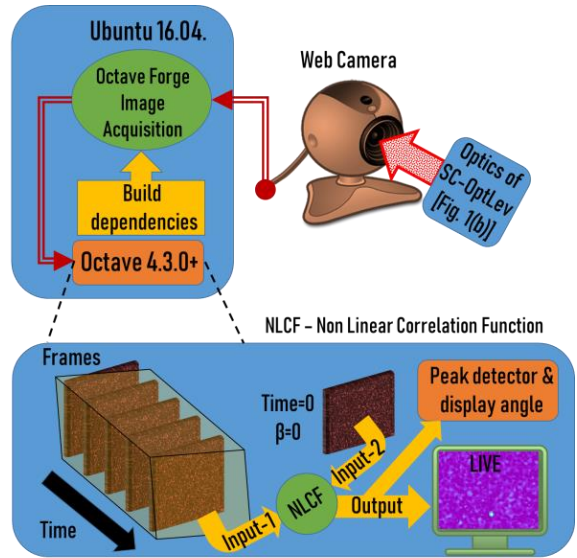

Fig. 3 Schematic of the computational module and the process sequence.

## 5. Summary and Conclusion

A SC-OptLev is designed and experimentally demonstrated to exhibit a dynamic range which is twice that of the conventional OptLev [25]. The feature of the extended dynamic range is due to the ability to detect a partial signature of the speckle pattern even when most of this pattern is out of the detection plane. However, we find experimentally that a conventional cross-correlation is not enough to detect the partial signature and among the five tested methods the non-linear correlation yielded the best performances. A computational module based on a completely open source GNU Linux-GNU Octave is presented with assistive computational codes to use a low-cost web camera for scientific application. The presented codes are not limited to SC-OptLev but can be adapted for many digital holography applications, especially for correlation optical systems [26-28], where real-time processing of holograms and simultaneous monitoring of many planes of an object are required.

### Acknowledgement

The authors thank LIGO R&D for India, IIT Madras and Inter-University Centre for Astronomy and Astrophysics, India for funding this research. For questions regarding installation of Ubuntu, building dependencies, please write to *muthaiahm@ee.iitm.ac.in*. This study was done during a research stay of JR at the Alfried Krupp Wissenschaftskolleg Greifswald.

# Authors' Biography

### A. Vijayakumar

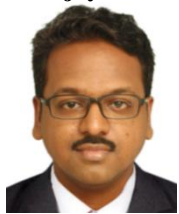

A.Vijayakumar has obtained his B.Sc and M.Sc from The American College, India in the year 2004 and 2007 respectively. He has obtained his M.Tech from Anna Univeristy, India in 2009. He obtained his Ph.D in Optics from Department of Electrical Engineering, Indian Institute of Technology Madras (IIT-M), India in 2015. He was a post-doctoral fellow in the Electro Optics research group, Department of Electrical and Computer Engineering, Ben Gurion University of the Negev, Israel from 2015-2018. He is currently working as a project officer in IIT-M. He is a member of OSA.

### D. Jayavel

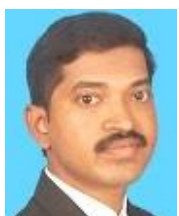

D. Jayavel has obtained his Diploma in Electronics and Communication Engineering from the Directorate of Technical Education in 2005 and currently pursing Bachelor of Engineering in Electronics and Communication Engineering in Anna University. He joined the Photonics Group, Indian Institute of Technology Madras as a Project Technician in 2006. In 2010, he became Junior Technician and from 2015 he is a Junior Technical Superintendent with the same group. He has been associated with various research projects such as design of Optical Time Domain Reflectormetry, Fiber Lasers, Fiber Bragg Gratings fabrication and sensors, free space optical communication and development of assistive technologies for differently-abled persons.

### M. Muthaiah

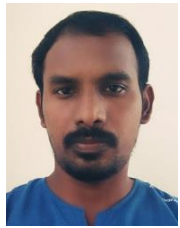

M. Muthaiah obtained his B.Tech in Electronics and Communication Engineering from Madras Institute of Technology Campus, Anna University in 2013. He is currently a Project Associate with the Department of Electrical Engineering, Indian Institute of Technology Madras, India. He has been working for the Center for NEMS and Nanophotonics on software development in open-source platform such as Linux. He was also working on site reliability engineering, and security.

### Shanti Bhattacharya

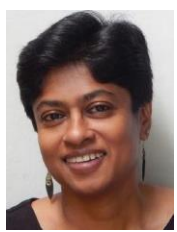

Shanti Bhattacharya completed her B.Sc Physics from WCC, Madras in 1990. She completed her M.Sc from Indian Institute of Technology Madras in 1992 and Ph.D from the same University on the topic of Optical Array Illuminators. She was awarded the Alexander von Humboldt award in 1998 at the [Technical University of Darmstadt, Germany](). After spending almost three years in Germany, she moved to Cambridge, USA where she worked at the MEMS division of Analog Devices. In 2002, she returned to India and worked part-time at the [Centre for Intelligent Optical Networks](). She joined the TeNeT group as an assistant professor in 2005. She is currently a Professor in the same institute. Her research interests include diffractive optics, optical coherence tomography, microfabrication, fabrication of diffractive optics on the fibre tip, Meta optics and MEMS.

**Joseph Rosen**

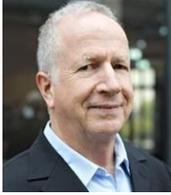

Joseph Rosen received the B.Sc., M.Sc., and D.Sc. degrees in Electrical Engineering from the Technion - Israel Institute of Technology, Haifa, Israel, in 1984, 1987, and 1992, respectively. He is a Benjamin. H. Swig Professor of Opto-electronics with the Department of Electrical and Computer Engineering, Ben-Gurion University of the Negev, Beer-Sheva, Israel. He has co-authored about 200 scientific journal papers, book chapters, and conference publications. His research interests include digital holography, optical microscopy, diffractive optics, statistical optics, biomedical optics, optical computing, and image processing. Dr. Rosen is a Fellow of the Optical Society of America (OSA) and the International Society for Optical Engineering (SPIE).